\documentclass[a4paper,12pt]{article}

\usepackage{mathptmx}
\usepackage{setspace}
\usepackage[left=2.5cm,right=2.5cm,top=2.5cm,bottom=2.5cm]{geometry}

\usepackage{titlesec}
\titleformat{\section}{\bfseries\large}{\thesection\ }{1em}{}
\titleformat{\subsection}{\bfseries}{\thesubsection\ }{1em}{}

\usepackage[hidelinks]{hyperref}

\usepackage{datetime}
\newdateformat{mydate}{\THEYEAR-\twodigit{\THEMONTH}-\twodigit{\THEDAY}}

\usepackage[sort&compress,numbers]{natbib}
\usepackage{abstract}
\usepackage{titling}
\usepackage{graphicx}
\usepackage{subfig}

\def \jpsi {J/\psi}
\def \to   {\rightarrow}
\def \ds   {D_s}
\def \dsm  {D_s^-}
\def \dsp  {D_s^+}
\def \dzb  {\bar{D}^0}
\def \dz   {D^0}
\def \cc   {+c.c.}

\def \gev  {~\mbox{GeV}}

\def \gevcc{~\mbox{GeV/$c^2$}}

\def \bb   {\mathcal{B}}
\def \km   {K^-}
\def \pip  {\pi^+}
\def \pim  {\pi^-}
\def \piz  {\pi^0}

\begin{document}

% Title
\begin{center}
    \vspace{2cm}
    {\huge\bfseries Search for charm rare decays at BESIII\par}
    \vspace{0.5cm}
    {\Large (Presented at the 32nd International Symposium on Lepton Photon Interactions at High Energies\par} 
    {\Large Madison, Wisconsin, USA, August 25-29, 2025) \par}
    \vspace{0.3cm}
    {\large Tianzi Song$^{a}$\footnote{Speaker. Email: songtz@mail2.sysu.edu.cn}, Zhengyun You$^{a}$ on behalf of BESIII Collaboration \par}
    \vspace{0.2cm}
    {\small
    $^a$ School of Physics, Sun Yat-sen University, Guangzhou, 510275, China \\
    }
    \vspace{0.3cm}
    {\mydate\today}
\end{center}

\vspace{0.4cm}

% Abstract
\begin{abstract}
\noindent
The BESIII experiment has collected 2.6 billion $\psi(3686)$ events, 10 billion  events, 20 $fb^{-1}$ of $D$ meson pairs at 3.773 GeV, and 7.33 $fb^{-1}$ of $D_sD_s^*$ events from 4.128 to 4.226 GeV. 
These huge data samples allow us to search for rare or forbidden processes in charm hadron decays. 
We summarize the recent research of charm rare decays at BESIII in this paper. 
% We report the search of the FCNC decays $J/\psi\rightarrow D^0\mu^+\mu^-$ and $D^+_s\rightarrow h(h')e^+e^-$. 
% The searches for $J/\psi$ weak decays containing a D meson and for $J/\psi\rightarrow \gamma D^0$ will also be presented. 
% In addition, we will introduce the search for baryon number violation via $\Lambda - \bar{\Lambda}$ oscillation in $J/\psi \rightarrow \Lambda \bar{\Lambda}$ decay, and the search for lepton number violation processes $D_s^+\rightarrow h^-h^0e^+e^-$ and $\phi\rightarrow\pi^+\pi^+e^+e^-$.
\end{abstract}

\vspace{0.6cm}

% Main Body
\section{Introduction}
The Standard Model~(SM) of particle physics has achieved remarkable success, yet it remains fundamentally incomplete, leaving critical questions like Dark Matter, matter-antimatter asymmetry, and the origin of neutrino masses unanswered. 
These limitations strongly suggest the existence of New Physics~(NP) beyond the SM's framework.
The search for NP at the intensity frontier often focuses on rare and forbidden processes. 
Within the SM, transitions that involve charmonium weak decays, flavor-changing neutral currents~(FCNC) or violate fundamental symmetries like Lepton Number~(LNV) and Baryon Number~(BNV) are severely suppressed or strictly prohibited. 
Consequently, the charm quark sector
% , benefiting from the effective suppression of FCNC decays by the GIM mechanism, 
offers an ideal, clean environment for highly sensitive searches of these NP effects. 
Any observation of these decays, or precise measurements that significantly deviate from SM predictions, would thus constitute an unambiguous signal of NP contributions. 
% The community is actively pursuing enhanced sensitivity across various fronts, including 
%, but not limited to, 
% the acquisition of larger data samples and the application of advanced methodologies, such as visualization techniques, to optimize analysis efficiency~\cite{vis}.

Beijing Spectrometer III~(BESIII)\cite{BESIII, BESIII unity} experiment operates in the $\tau$-charm energy region, covering center-of-mass energies from 1.84 to 4.95$\gev$. 
At present, BESIII has accumulated a world-leading dataset, including 10 billions $\jpsi$ events, 2.7 billions $\psi (3686)$ events, 20 $fb^{-1}$ of data collected at 3.773$\gev$ and over 20 $fb^{-1}$ collected above 4.0$\gev$ energies~\cite{besiiidata}. 
These extensive and unique data samples provide an opportunity to probe NP via highly sensitive searches for charm rare decays. 
Additionally, BESIII continually enhances its physics analysis capabilities through the application of advanced tools, such as event display techniques~\cite{unity}, to fully exploit its data for New Physics exploration~\cite{vis}.

\section{Charm rare decay measurements at BESIII}
\subsection{Charmonium weak decays}
\begin{figure}[t]
  \centering 
  \subfloat[$\jpsi\to\dsm\pi^+(\rho^+)$]{
        \label{fig:1.1}
        \includegraphics[width=0.4\textwidth]{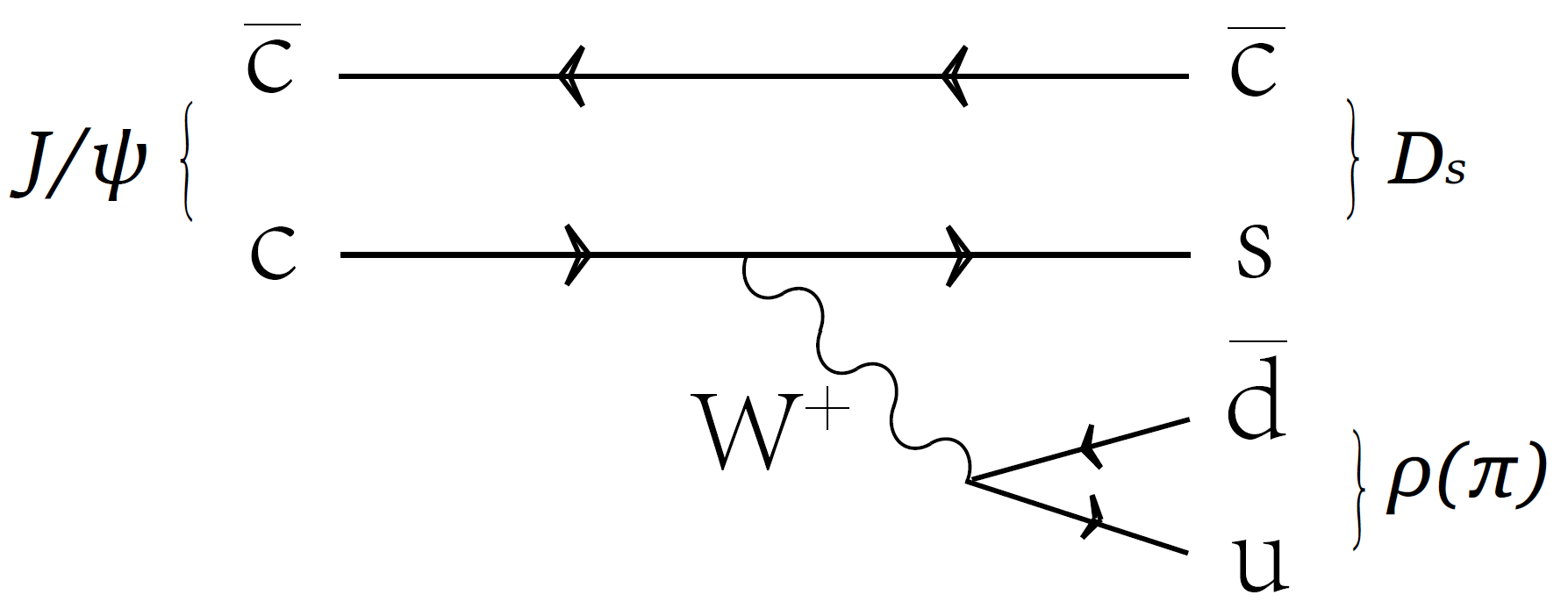}
  }
  % \hfill
  \subfloat[$\jpsi\to\dzb\pi^0(\rho^0/\eta)$]{
        % \label{fig:EicC}
        \includegraphics[width=0.26\textwidth]{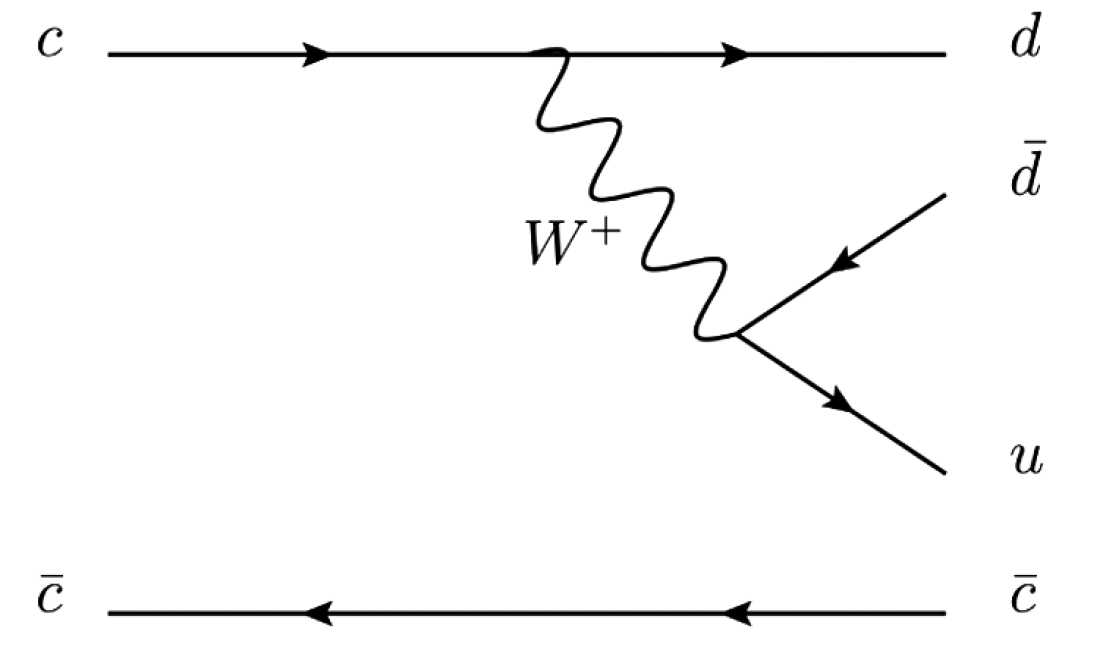}
  }
  \subfloat[$\jpsi\to D^-\pi^+(\rho^+)$]{
        % \label{}
        \includegraphics[width=0.26\textwidth]{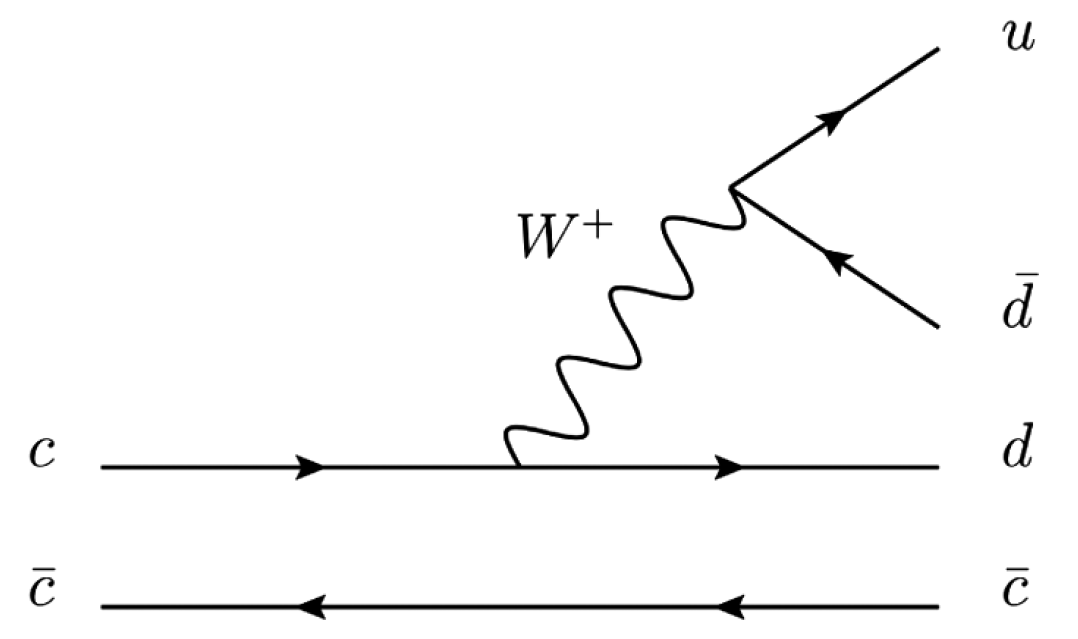}
  }
  \caption{Feynman diagram in tree level for charmonium weak decays.}
  \label{fig:charmonium}
\end{figure}

Since $\jpsi$ lies below the $D\bar{D}$ mass threshold, any decay mode involving a single charm meson must be mediated solely by the weak interaction with a light meson ($\jpsi\to D^{(0)}_{(s)}+$light meson). 
% it can only decay to a single charm meson $D_{(s)}^{(0)}$ accompanied with a light meson, such as $\pi$, $\rho$ or $\eta$. 
This kind of charmonium weak decays are highly suppressed within the SM, with the inclusive branching fraction~(BF) predictions around $10^{-8}$. 
% by strong and electromagnetic interaction, are rare but allowed in the SM. 
Various NP models beyond the SM, however, can significantly enhance this BF by up to three orders of magnitude, making these channels sensitive for NP model. Figure \ref{fig:charmonium} shows the Feynman diagram at the tree level for charmonium weak decays
$\jpsi\to\dsm\pi^+(\rho^+)$, $\jpsi\to\dzb\pi^0(\rho^0/\eta)$ and $\jpsi\to D^-\pi^+(\rho^+)$. 
Using the world largest sample of $(10087\pm44)\times 10^6\ \jpsi$ events collected at BESIII, we search for these charmonium rare weak decays~\cite{weak1, weak2}. 
No significant signal of these decays is found. 
The upper limits~(UL) of the BF at 90\% confidence level~(CL) are set as: $\bb(\jpsi\to\ds\rho^+)<8.0\times10^{-7}$, $\bb(\jpsi\to\ds\pi^+)<4.1\times10^{-7}$, $\bb(\jpsi\to\dzb\pi^0)<4.7\times10^{-7}$, $\bb(\jpsi\to\dzb\eta)<6.8\times10^{-7}$, $\bb(\jpsi\to\dzb\rho^0)<5.2\times10^{-7}$, $\bb(\jpsi\to D^-\pi^+)<7.0\times10^{-7}$, and $\bb(\jpsi\to D^-\rho^+)<6.0\times10^{-7}$. 
These ULs of BF for $\jpsi\to\dsm\pi^+(\rho^+)$ and $\jpsi\to\dzb\pi^+$ represent an improvement of $10^{-1}\sim10^{-3}$,
% one to three orders of magnitude, 
while the remaining channels are reported for the first time. 

%%%%%%%%%%%%%%%%%%%%%%%%%%%%%%%%%%%%%%%%%%%%%%%%%%%%
%%%%%%%%%%%%%%%%%%%%%%%%%%%%%%%%%%%%%%%%%%%%%%%%%%%%%

\subsection{FCNC decays}
% Compared to Flavor-chaning charged currents, 
In the SM, FCNC processes, such as $c\to ul^+l^-$, are strictly forbidden at the tree level.
% due to the absence of Z boson or photon coupling to different quark flavors.  
They can only proceed at the loop level, where the SM predictions are strongly suppressed by the GIM mechanism.
This suppression is particularly effective in the charm sector, leading to extremely low BFs within the SM, typically below $10^{-9}$. 
This high level of suppression makes charm FCNC decays a higly sensitive probe for NP, as any signal above the SM prediction would indicate NP contributions. 

\begin{figure}[t]
  \centering 
  \subfloat[$\jpsi\to D^0l^+l^-$]{
        \label{fig:1.1}
        \includegraphics[width=0.4\textwidth]{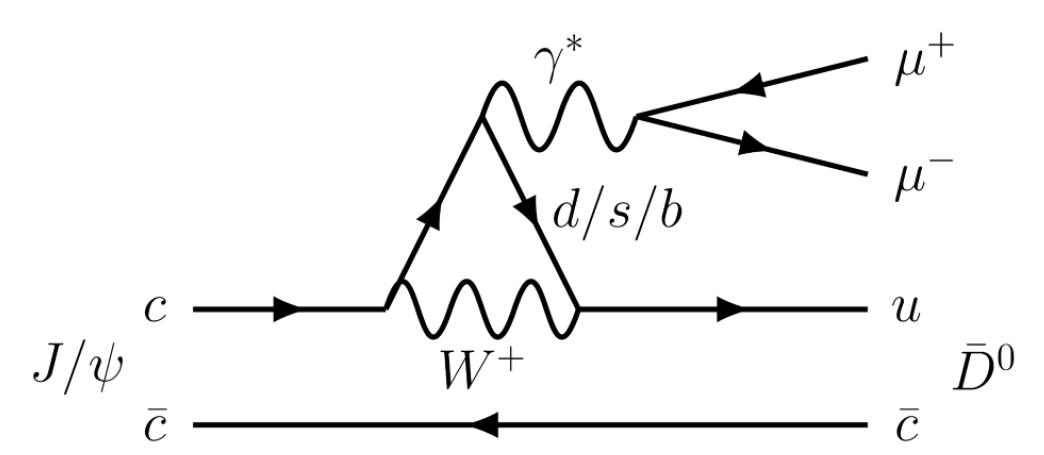}
  }
  % \hfill
  \subfloat[$\jpsi\to\gamma D^0$]{
        % \label{fig:EicC}
        \includegraphics[width=0.4\textwidth]{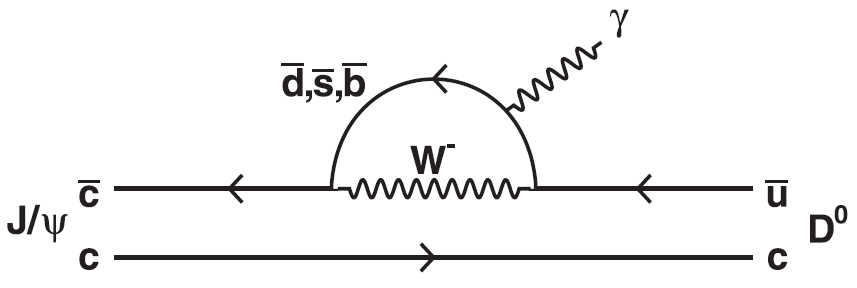}
  }\\
  % \hfill
  \subfloat[The SD contributions in $D_{(s)}\to h(h')l^+l^-$ processes. ]{
        % \label{}
        \includegraphics[width=0.57\textwidth]{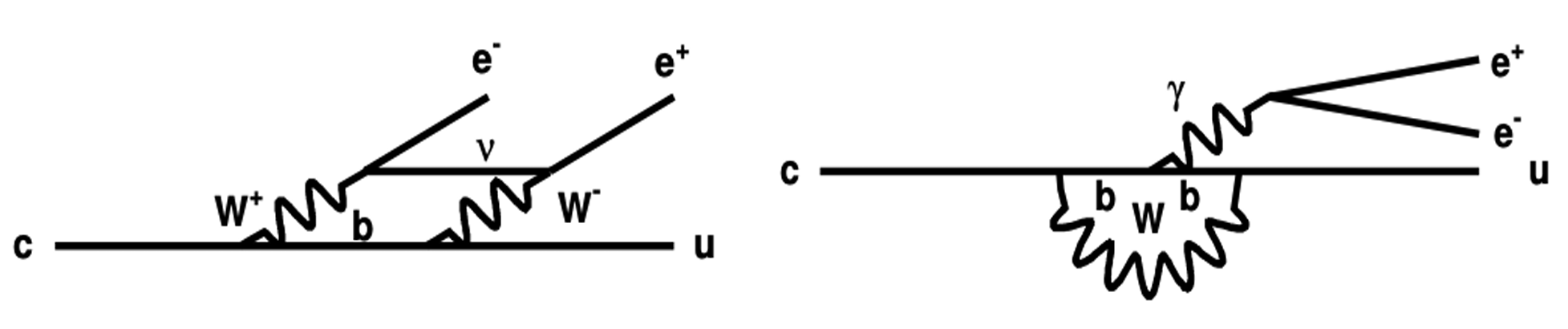}
  }\hfill
  % \subfloat[$\jpsi\to D^-\pi^+(\rho^+)$]{
  %       % \label{}
  %       \includegraphics[width=0.28\textwidth]{2.4.png}
  % }\hfill
  \subfloat[The LD contributions, $D_s^+\to\pi^+(\rho^+)\phi$. ]{
        % \label{}
        \includegraphics[width=0.38\textwidth]{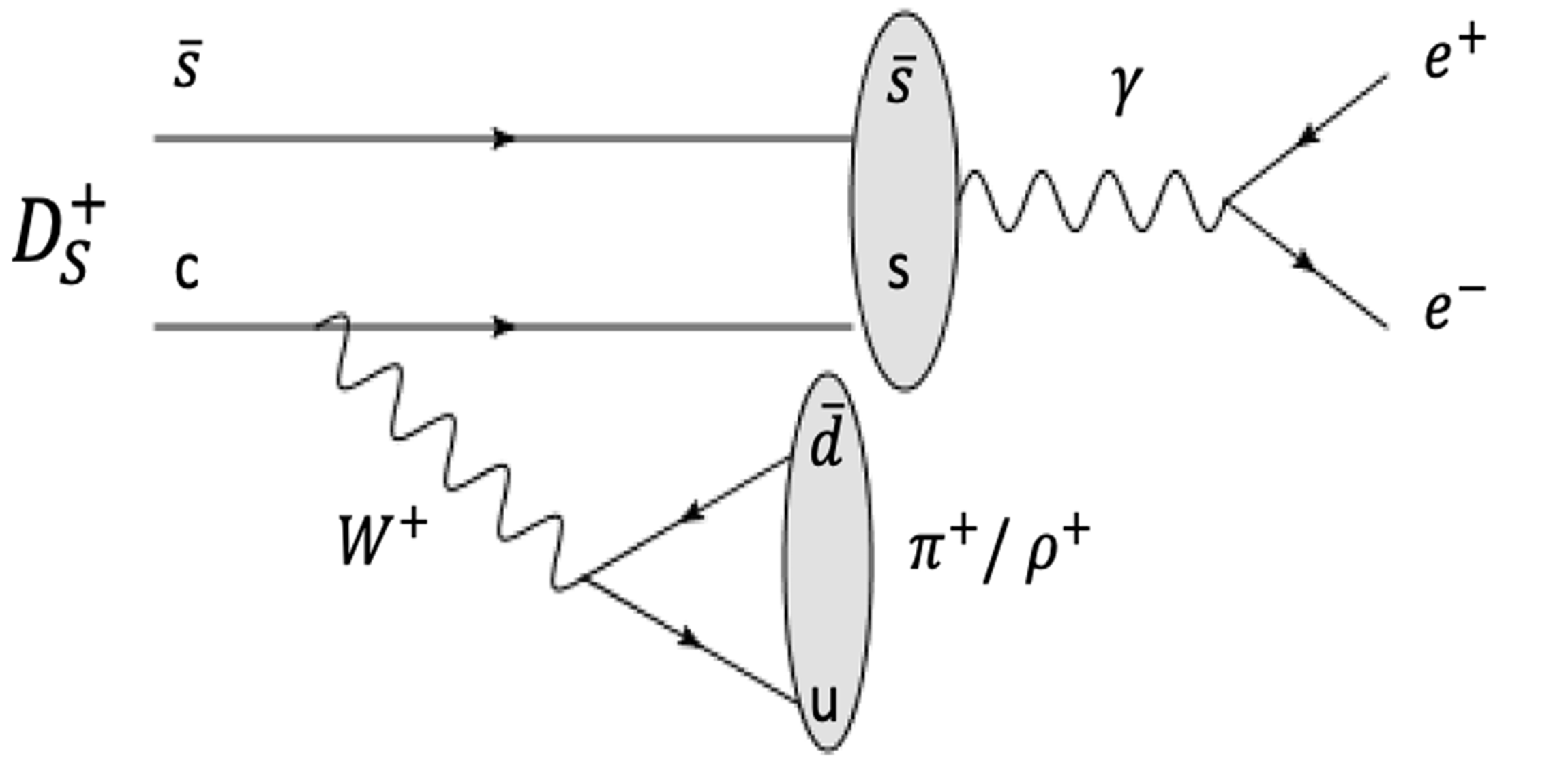}
  }
  
  \caption{Feynman diagram for FCNC processes and one of the possible non-FCNC contribution.}
  \label{fig:FCNC}
\end{figure}

Besides the $J/\psi$ weak decays to a single charm meson and a light meson, it is also possible for the $J/\psi$ to decay into a single charm meson accompanied by a lepton pair,
% (e.g., $D^{(0)}_{(s)} l^+ l^-$), 
albeit with an extremely low BF. 
Furthermore, if we consider one fewer decay vertex, the $\jpsi$ can decay directly into a single charm meson and a photon $\jpsi\to\gamma\dz$, which is expected to have a slightly enhanced BF compared to the weak decay $\jpsi\to\dz\mu^+\mu^-$. 
The Feynman diagrams for these decays are shown in Figure \ref{fig:FCNC} (a) and (b). 
Searches for $\jpsi\to D^0\mu^+\mu^-$ and $\jpsi\to\gamma D^0\cc$ have been performed using $(10087\pm44)\times10^6\ \jpsi$ events~\cite{fcnc1, fcnc2}. 
Three decay modes $\dz\to\km\pip$, $\dz\to\km\pip\piz$, and $\dz\to\km\pip\pip\pim$ are used to reconstruct $\dz$ meson. 
The observed data was consistent with a background-only hypothesis in both decays. 
Therefore, we set the following UL at the 90\% confidence level:  $\bb(\jpsi\to\dz\mu^+\mu^-)<1.1\times10^{-7}$ and $\bb(\jpsi\to\gamma\dz)<9.1\times10^{-8}$. 
Both results constitute the most stringent ULs to date for their respective decays. 
Notably, the search for $\jpsi\to\dz\mu^+\mu^-$ represents the first search for a charmonium FCNC process involving muons in the final state. 
%%%%%%%%%%%%%%%%%%%%%%%%%%%%%%%%%%%%%%%%%%%%%%%%%%%%
%%%%%%%%%%%%%%%%%%%%%%%%%%%%%%%%%%%%%%%%%%%%%%%%%%%%%

A series of processes $D_s^+\to h(h')e^+e^-$ are studied with 2.93 $fb^{-1}$ data samples at the center-of-mass energy $\sqrt{s}=3.773\gev$~\cite{fcnc3}, where $h$ denotes hadron such as kaon or pion. 
These decays include short-distance~(SD) contributions (via the $c\to ul^+l^-$ FCNC transition) and long-distance~(LD) contributions, which are dominated by non-FCNC processes, shown in Figure \ref{fig:FCNC} (c) and (d).
Because the BFs for the SD-only FCNC decays are predicted to be $\sim 10^{-9}$, they are often overshadowed by much larger LD contributions (typically $\sim 10^{-6}$), making these searches exceptionally challenging to observe. 
As a result, we observe the LD processes $\dsp\to\pip\phi$, $\phi\to e^+e^-$ with a statistical significance $7.8\ \sigma$, and $\dsp\to\rho^+\phi$, $\phi\to e^+e^-$ with a statistical significance $4.4\ \sigma$. 
These LD BFs are agree with the CLEO result, and are also consistent with the products of the PDG values $\bb(\dsp\to\pip\phi)\cdot\bb(\phi\to e^+e^-)$ and $\bb(\dsp\to\rho^+\phi)\cdot\bb(\phi\to e^+e^-)$ within uncertainties. 
No significant signal of the SD-only FCNC processes is observed, so the ULs of BF are set at 90\% confidence level. 
These results represent the first ULs on the BFs of these decays. 

\subsection{LNV decays}
\begin{figure}[ht]
  \centering 
  \includegraphics[width=0.8\textwidth]{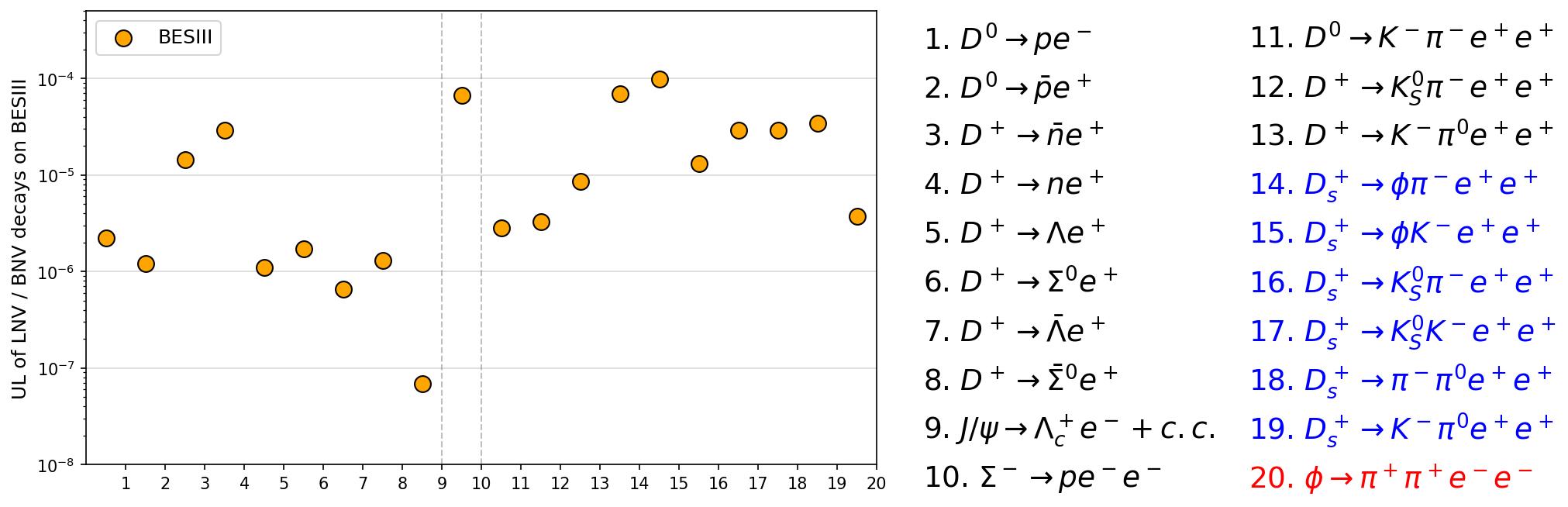}
  \caption{The ULs on the BFs for LNV and BNV processes on BESIII. (Color online) }
  \label{fig:LNVBNV}
\end{figure}

In the SM, 
% the total lepton number is conserved in all interactions, and 
the fundamental neutrinos are massless. 
However, the observation of neutrino oscillations provides strong evidence that neutrinos have non-zero mass, necessitating an extension to the SM. 
Models explaining neutrino mass, such as See-saw Mechanism~\cite{seesaw}, proposes the neutrinos are Majorana neutrinos, which can be manifested through LNV processes with $\Delta L=2$. 
Figure \ref{fig:LNVBNV} shows the summary of BESIII recent result for LNV and BNV processes.
% with $\Delta L =2$
%%%%%%%%%%%%%%%%%%%%%%%%%%%%%%%%%%%%%%%%%%%%%%%%%%%%
%%%%%%%%%%%%%%%%%%%%%%%%%%%%%%%%%%%%%%%%%%%%%%%%%%%%%

A search for LNV decays $\dsp\to h^-h^0e^+e^+$ was performed using $7.33\ fb^{-1}$ of $\ds^{*\pm}\ds^{\mp}$ data sample collected at center-of-mass energies ranging from 4.128 to 4.226$\gev$~\cite{lnv1}. 
% This analysis content six decay channels: the Cabibbo Favor decays $\dsp\to\phi\pim e^+e^+$, the Singly Cabibbo Suppressed decays $\dsp\to\phi K^- e^+e^+$, $\dsp\to\ksz\pim e^+e^+$, the Doubly Cabibbo Suppressed decays $\dsp\to\ksz\km e^+e^+$, the W-exchange process $\dsp\to\pim\piz e^+e^+$, and $\dsp\to\km\piz e^+e^+$. 
No obvious signal events was observed in the signal region for any of channels. 
Consequently, we set the ULs on the BFs at 90\% confidence level. 
The results of this anaylsis are denoted by blue labels in Figure \ref{fig:LNVBNV}. 
This work constitutes the first measurement of the four-body $\Delta L=2\ \dsp$ decays.
% , establishing the most stringent limits to date on these six channels.
Additionally, the search for a Majorana neutrino is also performed in this analysis via  $\dsp\to\phi e^+\nu_m(\to\pim e^+)$.
We investigated the results across a range of assumed masses, $m_{\nu_m}$, from 0.20 to 0.80$\gevcc$, and the ULs on the BFs at 90\% confidence level are determined as a function of $m_{\nu_m}$, ranging from around $10^{-5}\sim10^{-2}$.

The LNV decay $\phi\to\pip\pip e^-e^-$ is searched for via $\jpsi\to\eta\phi$ using 10 billions $\jpsi$ events collected at BESIII~\cite{lnv2}. 
To get rid of the large uncertainty from $\bb(\jpsi\to\phi\eta)$, the reference channel $\phi\to K^+K^-$ is analyzed. 
% In this situation, the BF for the signal decay can be calculated as:
% \begin{equation}
%     \bb(\phi\to\pip\pip e^-e^-)=\bb(\phi\to K^+K^-)\times\frac{N_{\pip\pip e^+e^+}/\varepsilon_{\pip\pip e^-e^-}}{N_{K^+K^-}/\varepsilon_{K^+K^-}}
% \end{equation}
% $\bb(\phi\to\pip\pip e^-e^-)=\bb(\phi\to K^+K^-)\times\frac{N_{\pip\pip e^+e^+}/\varepsilon_{\pip\pip e^-e^-}}{N_{K^+K^-}/\varepsilon_{K^+K^-}}$. 
No significant signal event is found in the signal region, so we set the UL on the BF at 90\% confidence level to be $1.3\times 10^{-5}$. 
This result is presented by red label in Figure \ref{fig:LNVBNV}. 

\subsection{BNV decays}
To explain mater-antimatter asymmetry, Sakharov proposed three conditions, one of which is the baryon number violation. 
In Grand Unified Theory~(GUT), processes such as proton decay $p\to e^+\piz$ are allowed, which simultaneously breaks BN and LN while conserving $\Delta (B-L)$~\cite{seesaw}. 
Moreover, a Majorana particles with small masses would imply the presence of $\Delta (B-L)=2$ interactions~\cite{bl2}, thereby suggesting the existence of nucleon-antinucleon ($n-\bar{n}$) oscillation. 
To date, many experiments have studied about the $N-\bar{N}$ oscillation, while few results have been reported for hyperons oscillation. 
Using the world's largest data sample of 10 billions $\jpsi$ events, the BESIII experiment performed the first search for $\Lambda-\bar{\Lambda}$ oscillation via $\jpsi\to\Lambda\Lambda\cc$, yielding no evidence~\cite{bnv1}. 
Consequently, ULs at 90\% confidence level are established for the time-integrating probability of $\Lambda-\bar{\Lambda}$ oscillation $P(\Lambda)<1.4\times 10^{-6}$ and the oscillation parameter $\delta(m_{\Lambda\bar{\Lambda}})<2.1\times10^{-18}$. 
This is the most stringent constraint in the world on corresponding parameters for  $\Lambda-\bar{\Lambda}$ oscillation. 

\section{Summary}
In summary, we have presented a comprehensive overview of the recent searches for rare or even forbidden process in the charm sector, utilizing the world-leading data samples collected by the BESIII experiment. 
In this paper, we presented some recent results searching for charm rare decays on BESIII including charmonium weak decays,
% with a single $D_{(s)}^{(0)}$ meson, 
Flavor Changing Neutral Currents, Lepton Number Violation, and Baryon Number Violation. 
% The exceptional sensitivity provided by the 10 billions $\jpsi$ and extensive data samples collected at other crucial energies, offers us to sufficient oppotunities to explore a wide range of NP scenarios. 
With the continued data-taking process, BESIII will have the opportunity to perform more precise searches for NP models. 
We highly anticipate that more excellent results will be published in the near future.

\section*{Acknowledgements}
This work is supported by the National Key R\&D Program of China under Contracts Nos. 2023YFA1606000; National Natural Science Foundation of China (Grant No. 12175321, U1932101, 11975021, 11675275).

% References

\end{document}